\documentclass[reprint,pre,floats,aps,amsmath,amssymb,notitlepage]{revtex4-2}

\usepackage{physics,graphicx,xcolor,hyperref}
\usepackage{colortbl}

\usepackage{dsfont}
\usepackage{subcaption}
\usepackage{enumerate}
\usepackage{mathrsfs}
\usepackage{amsmath}
\usepackage{ragged2e} 
\usepackage{caption}
\newcommand{\bH}{{\bf H}}
\newcommand{\bA}{{\bf A}}
\newcommand{\bB}{{\bf B}}
\newcommand{\bG}{{\bf G}}
\newcommand{\bJ}{{\bf J}}
\newcommand{\bh}{{\bf h}}
\newcommand{\bt}{{\bf t}}
\newcommand{\bSg}{{\bf \Sigma}}
\newcommand{\bPs}{\boldsymbol{\Psi}}
\newcommand{\bmu}{\boldsymbol{\mu}}
\newcommand{\bnu}{\boldsymbol{\nu}}

\newcommand{\bet}{\boldsymbol{\eta}}
\newcommand{\bzet}{\boldsymbol{\zeta}}

\newcommand{\bK}{{\bf K}}
\newcommand{\bQ}{{\bf Q}}
\newcommand{\bR}{{\bf R}}
\newcommand{\bS}{{\bf S}}

\newcommand{\bscrT}{{\pmb{\mathscr{T}}}}
\newcommand{\bU}{{\bf U}}
\newcommand{\bW}{{\bf W}}

\begin{document}

\title{Spectral statistics of interpolating random circulant matrix and its applications to random circulant graphs}

\author{Sunidhi Sen}
\email{sensunidhi96@gmail.com}
\affiliation{%
Department of Physics, Shiv Nadar Institution of Eminence,
Gautam Buddha Nagar, Uttar Pradesh 201314, India
}%
\author{Himanshu Shekhar}
\email{himanshuphy02@gmail.com}
\affiliation{%
Department of Physics, Shiv Nadar Institution of Eminence,
Gautam Buddha Nagar, Uttar Pradesh 201314, India
}%
\author{Santosh Kumar}
\email{Deceased}
\affiliation{%
Department of Physics, Shiv Nadar Institution of Eminence,
Gautam Buddha Nagar, Uttar Pradesh 201314, India
}%


\begin{abstract}
We consider a versatile matrix model of the form ${\bf A}+i {\bf B}$, where ${\bf A}$ and ${\bf B}$ are real random circulant matrices with independent but, in general, non-identically distributed Gaussian entries. For this model, we derive exact results for the joint probability density function and find that it is a multivariate Gaussian. Arbitrary order marginal density therefore also readily follows. It is demonstrated that by adjusting the averages and variances of the Gaussian elements of ${\bf A}$ and ${\bf B}$, we can interpolate between a remarkably wide range of eigenvalue distributions in the complex plane. In particular, we can examine the crossover between a random real circulant matrix and a random complex circulant matrix. We also extend our study to include Wigner-like and Wishart-like matrices constructed from our general random circulant matrix. To validate our analytical findings, Monte Carlo simulations are conducted, which confirm the accuracy of our results. Additionally, we compare our analytical results with the spectra of adjacency matrices from various random circulant graphs. Despite the difference in entry distributions--Gaussian in our model and non-Gaussian in the adjacency matrices--the densities show excellent agreement in the large-dimension limit.
\end{abstract}

\maketitle

\section{Introduction}
\label{SecIntro}

Circulant matrices, a special type of Toeplitz matrices, exhibit a distinct pattern in which each row is formed by cyclically shifting the entries of the preceding row. This inherent symmetry and algebraic structure make circulant matrices immensely valuable across various domains of mathematics, physics, engineering, and computer science~\cite{Davis1979,Gray2006}. They offer efficient solutions for a wide range of problems including difference and differential equations~\cite{BGN1970,Wilde1983,Gilmour1988,DRR2003,CLC2013}, graph theory~\cite{ET1970,Vilfred2004,AP1979,MM1997,RR2018}, time-series analysis~\cite{Pollock2002,FY2003,BD2006}, signal and image processing~\cite{BLM1990,Andrecut2008,WWR2008,TDB2015}, computer vision~\cite{HCMB2012,DQT2017}, cryptography~\cite{DR2002}, coding theory~\cite{MS1977}, vibrational analysis~\cite{OSSPP2014}, statistical physics~\cite{BK1952}, quantum mechanics~\cite{ST1976,Aldrovandi2001}, among others.  
 
Random variants of circulant matrices have also received attention due to their natural occurrence in problems related to random walks~\cite{Aldrovandi2001}, stochastic time series analysis~\cite{Pollock2002}, 5G communication schemes~\cite{TDB2015}, and so on. Moreover, there is a natural curiosity regarding the behavior of their spectra compared to the classical random matrices such as Wigner and Wishart~\cite{Mehta2004,Forrester2010,Wigner1955,Wigner1957,Wigner1958,AGZ2009,Wishart1928,MP1967,GMW98,ABF18}, as well as non-Hermitian Ginibre matrices~\cite{Ginibre1965, FKS1997, BF2024, ADM2023,Molag2023,SK2024,ABK2021,BBD2023}. Notably, in~\cite{JS2008,SJ2013,AS2022}, the local spectral fluctuations of random circulant matrices have been studied using nearest neighbor distribution. Additionally, several interesting results, including those pertaining to limiting spectral density and associated moments, can be found in~\cite{Meckes2009,BS2018,Bose2018}. However, despite these significant contributions, there still remains much to be explored, particularly in terms of exact results. The present work is an attempt to contribute in this direction.
 
We consider a very general $N$-dimensional complex circulant random matrix, ${\bf H}={\bf A}+i{\bf B}$, where ${\bf A}$ and ${\bf B}$ are real circulant  matrices, whose entries have been chosen as Gaussian variables which are independent but in general taken from non-identical distributions. For this random matrix model, we derive an exact closed-form expression for the joint probability density function (JPDF) of all eigenvalues, revealing that it follows a multivariate Gaussian distribution. Consequently, we can readily obtain the exact marginal densities for any subset of eigenvalues. We show that the eigenvalues can exhibit a vast range of behavior in the complex plane as one tunes averages and variances of the Gaussian elements of ${\bf A}$ and ${\bf B}$. In particular, the interpolation between real circulant and complex circulant random matrix can be realized. Finally, we also examine the eigenvalues of Wigner-like and Wishart-like matrices constructed out of ${\bf H}$. We also carry out Monte Carlo simulations and find the results thereof to be consistent with our analytical results.

The remainder of the paper is organised as follows. In Sec.~\ref{SecMatMod}, we introduce our random circulant matrix model and derive the JPDF of real and imaginary parts of all eigenvalues. The expressions of various marginal densities are also provided in this section. Afterwards, in Secs.~\ref{SecWig} and~\ref{SecWis}, we study Wigner-like and Wishart-like matrices constructed out of the random circulant matrix ${\bf H}$. In Sec.~\ref{Application}, we apply our analytical results in the study of spectra of adjacency matrices from various random circulant graphs. We conclude with a summary of our results along with discussion of some possible future directions in Sec.~\ref{SecSum}.

\section{A general random circulant matrix model}
\label{SecMatMod}

 We consider the $N$-dimensional random circulant matrix, 
 \begin{equation}
 \label{matmod}
 \bH=\bA+i\bB,
 \end{equation} 
where $\bf A$ and $\bf B$ are two real circulant matrices. The complex circulant matrix $\bf H$, therefore, possesses the structure, 
\begin{equation}
   \bf{H}=\begin{bmatrix}
  h_1 & h_N & h_{N-1} &\ldots & h_2\\
  h_2 & h_1 & h_N & \ldots & h_3\\
  h_3 & h_2 & h_1 & \ldots & h_4\\
  \vdots & \vdots & \vdots & \ddots & \vdots\\
  h_N & h_{N-1} & h_{N-2} & \ldots & h_1
   \end{bmatrix}, 
\end{equation}   
with  $h_r=(a_r+i b_r)$; $r= 1,\dots,N$. Specifically, each element is given by $h_{p,q}=h_{((p-q) \mathrm{mod} N)+1}$, where $p,q=1,\ldots,N$ are the row and column indices. This relationship guarantees that the same element appears along every diagonal, emphasizing the cyclic structure of $\bH$. The elements $a_j$ and $b_j$, ($j=1,...,N$), are taken to be independent but non-identical Gaussians, with corresponding averages and variances as $u_j, v_j$ and $\sigma_j^2,\tau_j^2$, respectively. Therefore, they are distributed as
   \begin{equation}
P_{\boldsymbol{a}}(\{a\})=\prod_{j=1}^{N} \frac{1}{(2 \pi \sigma_j^2)^{1/2}} \exp{\left(-\frac{(a_j-u_j)^2}{2 \sigma_j^2}\right)},
\end{equation}
\begin{equation}
P_{\boldsymbol{b}}(\{b\})=\prod_{j=1}^N \frac{1}{(2\pi \tau_j^2)^{1/2}} \exp{\left(-\frac{(b_j-v_j)^2}{2 \tau_j^2}\right)}.
\end{equation}   
Observe that the matrix model $\alpha \bA+\beta \bB$, where $\alpha,\beta$ are real, can be mapped to the above model since $\alpha$ and $\beta$ can be absorbed within the averages $u_j,v_j$ and variances $\sigma_j^2, \tau_j^2$, yielding the modified averages and variances as $\alpha u_j, \beta v_j$ and $\alpha^2\sigma_j^2, \beta^2\tau_j^2$, respectively. We further observe that in the limit $v_j,\tau_j\to 0$ for $j=1,...,N$, effectively only $\bA$ survives, i.e., $\bH$ is purely real. Similarly when $u_j,\sigma_j\to 0$ for $j=1,...,N$, only $\bB$ survives so that $\bH$ is purely imaginary. 
 
 We are interested in the distribution of eigenvalues of $\bH$ and their dynamics in the complex plane as the averages and variances of the matrix elements of $\bA$ and $\bB$ are varied. From the properties of circulant matrices, we know that the eigenvalues of the matrix $\bf{H}$ can be written as~\cite{Davis1979,Gray2006}, 
  \begin{equation}
  \label{lambdas}
    \lambda_j=\sum_{r=1}^N h_r\, \omega_j^{(N-r+1)};~~j=1,...,N,
 \end{equation} 
where $\omega_{j}$ are the $N$th roots of unity,
\begin{align}
\label{unityroots}
\omega_{j}=\exp[i2\pi (j-1)/N].
\end{align}
Moreover, the diagonalizing matrix $\bU$ for any circulant matrix is unitary. The elements of the matrix $\bU$ are given by 
\begin{equation}
U_{j,k}=\frac{1}{\sqrt{N}}\omega_j^{k-1};~~j,k=1,...,N.
\end{equation}
The matrix $\bU$, often referred to as the Fourier matrix, holds significant importance in various fields, including graph signal processing~\cite{OFKMV2018,Miegham2023}, quantum computing~\cite{WL2018,Parthasarathy2006}, and coding theory~\cite{Massey1998,YMA2017}.

Expressing $\lambda_j$ above in terms of the matrix elements of $\bA$ and $\bB$, and separating the real and imaginary parts, we obtain
\begin{align}
\label{eigenValH}
\lambda_j&=\sum_{r=1}^N (a_r C_{j,r}-b_r S_{j,r})+i\sum_{r=1}^N (a_r S_{j,r}+b_r C_{j,r}),
\end{align}
where
\begin{align}
C_{j,r}= \cos\left[\frac{2 \pi (j-1) (N-r+1)}{N}\right],\\
S_{j,r}=\sin\left[\frac{2 \pi (j-1) (N-r+1)}{N}\right].
\end{align}
These trigonometric functions arise naturally by rewriting Eq.~\eqref{unityroots} using Euler's formula. The complex eigenvalues 
$\lambda_j$ can be represented as components of a vector $\bet=(\eta_1,\eta_2,...,\eta_{2N-1},\eta_{2N})^T$, where $(\cdot)^T$ denotes the transpose operation. Specifically, we write
\begin{align}
\lambda_j\equiv \eta_{2j-1}+i \eta_{2j},
\end{align}
indicating that the odd-indexed components of 
$\bet$ correspond to the real parts of the $\lambda_j$'s, while the even-indexed components correspond to the imaginary parts, as in Eq.~\eqref{eigenValH}. This vector representation of eigenvalues combines the real and imaginary parts into a single structure, making it easier to handle real-valued vectors compared to directly working with complex numbers.

To obtain the joint distribution of the $\lambda$'s, or equivalently that of the $\eta$'s, we need to integrate over the Gaussian variables $a_j$ and $b_j$. For compactness, let us define the following quantities. We consider a $2N$-dimensional column vector $\bh$ comprising $a_j$ and $b_j$ as $\bh=(a_1, b_1,....a_N, b_N)^T$.
Clearly, $\bh$ is governed by the multivariate Gaussian probability density function,
\begin{align}
P_{\bh}  (\bh)=&\frac{1}{[(2 \pi)^{2 N}\det\bSg]^{1/2}}\exp\left[-\frac{1}{2} (\bh -\bmu)^T \bSg^{-1} (\bh -\bmu)\right].
\end{align}
Here, $\bmu=(u_1,v_1,...,u_N,v_N)^T$ is the mean vector and $\bSg=\mathrm{diag}(\sigma_1^2,\tau_1^2,...,\sigma_N^2,\tau_N^2)$ is the covariance matrix. Now, it can be verified that
\begin{equation}
\label{etas}
\eta_{2j-1}=\bh^T \bK_1 \bt_j,~~\eta_{2j}=\bh^T \bK_2 \bt_j,
\end{equation}
for $j=1,..,N$, where we have defined $\bK_1=\mathds{1}_N\otimes\sigma_z$ and $\bK_2=\mathds{1}_N\otimes\sigma_x$, with $\mathds{1}_N$ representing the $N$-dimensional identity matrix. The matrices $\sigma_z=\begin{pmatrix}1 & 0\\0 & -1\end{pmatrix}$ and $\sigma_x=\begin{pmatrix}0 & 1\\1 & 0\end{pmatrix}$ are two of the Pauli matrices. Also, $\bt_{j}=(C_{j,1},S_{j,1},...,C_{j,N},S_{j,N})^T$. The joint probability density function of the real and imaginary parts of eigenvalues of the matrix $\bH$ can now be obtained by integrating over $\bh$, viz.,
\begin{align}
\nonumber
& P(\eta_1,\eta_2,\ldots,\eta_{2 N -1},\eta_{2 N})\\
&~~=\int d \bh\, P_{\bf{ h}}(\bh)\prod_{j=1}^{N} \delta(\eta_{2 j-1}-\bh^T \bK_1 \bt_j)\, \delta(\eta_{2j}-\bh^T \bK_2\bt_j).
\end{align}
The $\bh$-integral can be performed using the characteristic function (CF) approach~\cite{Ushakov1999} by going to the Fourier space, $\{\eta_j\mapsto \zeta_j\}$, as this method effectively accounts for the possibility of $\eta_j$ taking negative values. The resulting CF is given by,
\begin{align}
\nonumber
&\Phi(\zeta_1,\zeta_2,\ldots,\zeta_{2 N -1},\zeta_{2 N}) \\ \nonumber& =\int \prod_{j=1}^{2N} d\eta_j \exp(i\zeta_j \eta_j)P(\eta_1,\eta_2,\ldots,\eta_{2 N -1},\eta_{2 N})\\
&=\int d \bh\, P_{\bf{ h}}(\bh)\exp\bigg[i\bh^T\sum_{j=1}^{N}\left(\bK_1 \bt_j \zeta_{2 j-1} + \bK_2 \bt_j\zeta_{2 j}\right)\bigg],
\end{align}
where we used Eq.~\eqref{etas} to replace $\eta_j$. Now, we note that $\sum_{j=1}^{N}\left(\bK_1 \bt_j \zeta_{2 j-1} + \bK_2 \bt_j\zeta_{2 j}\right)=\bQ \bzet$, where $\bQ=(
\bK_1 \bt_1,\bK_2 \bt_1, \bK_1 \bt_2,\bK_2 \bt_2,\ldots, \bK_1 \bt_N,\bK_2 \bt_N  )$ is a $2N\times 2N$-dimensional matrix and $\bzet=(\zeta_1, \zeta_2,...,\zeta_{2N-1},\zeta_{2N})^T$ is a $2N$-dimensional column vector. Therefore, the above equation can be rewritten as,
\begin{align}
\nonumber
&\Phi(\bzet)=\int d \bh\, P_{\bf{ h}}(\bh) \exp(i\bh^T\bQ\bzet)\\
\nonumber
&~~~~~~ = \int d \bh\,\frac{1}{[(2 \pi)^{2 N}\det\bSg]^{1/2}}\\
&~~~~~~\times \exp\left[-\frac{1}{2} (\bh -\bmu)^T \bSg^{-1} (\bh -\bmu)+i\bh^T\bQ\bzet\right].
\end{align}
The above multidimensional-Gaussian integral can be readily performed to yield
\begin{align}
\Phi(\bzet)= \exp\left(-\frac{1}{2}\bzet^T \bscrT \bzet+i\bnu^T \bzet \right),
\end{align}
where $\bnu=\bQ^T \bmu$ and $\bscrT=\bQ^T\bSg \bQ$. We now perform the inverse Fourier transform and return back to the $\bf{\eta}$ space, leading us to the desired expression,
\begin{align}
\label{mainjpd}
P(\bet)=&\frac{1}{[(2 \pi)^{2 N}\det\bscrT]^{1/2}}\exp\left[-\tfrac{1}{2} (\bet -\bnu)^T \bscrT^{-1} (\bet -\bnu)\right].
\end{align}
This equation gives the joint probability density function of real and imaginary parts of the \emph{ordered} eigenvalues of $\bH$, as in Eq.~\eqref{lambdas}. Evidently, it is a multivariate Gaussian distribution. An immediate consequence is that any marginal of the above is again a (multivariate)-Gaussian distribution. For instance, the joint distribution of $\eta_{j_1},...,\eta_{j_r}$, with $r\le 2N$ is given by 
 \begin{align}
 \label{marginal-joint}
\widetilde{P}(\widetilde{\bet})= \frac{1}{[(2 \pi)^{r}\det \widetilde{\bscrT}]^{1/2}} \times\exp\left[-\frac{1}{2} (\widetilde{\bet} -\widetilde{\bnu})^T \widetilde{\bscrT}^{-1} (\widetilde{\bet} -\widetilde{\bnu})\right],
\end{align}
where the quantities with tilde on top have been obtained by removing all variables/parameters having indices other than $j_1$ to $j_r$. In particular, the marginal density involving the real ($\eta_{2j-1}$) and imaginary $(\eta_{2j})$ parts of a specific $\lambda_j$ is given by Eq.~\eqref{marginal-joint} with
$\widetilde{\bet}=\begin{pmatrix} \eta_{2j-1} \\ \eta_{2j} \end{pmatrix}, \widetilde{\bnu}=\begin{pmatrix} \nu_{2j-1} \\ \nu_{2j} \end{pmatrix}, \widetilde{\bscrT}=\begin{pmatrix} \mathscr{T}_{2j-1,2j-1}  & \mathscr{T}_{2j-1,2j}\\ \mathscr{T}_{2j,2j-1} & \mathscr{T}_{2j,2j}\end{pmatrix}$. Notably, the marginal density of a single component $\eta_j$ is just a Gaussian distribution, 
\begin{align}
\label{md1}
p(\eta_j)=\frac{1}{(2\pi \mathscr{T}_{jj})^{1/2}} \exp\left[-\frac{(\eta_j-\nu_j)^2}{2\mathscr{T}_{jj}}\right].
\end{align}
The above expressions give the distribution of the eigenvalue(s) ordered according to the ordering of the $N$th roots of unity, as in Eq.~\eqref{unityroots}. The joint distribution of unordered $\eta_j$ can be obtained by symmetrizing Eq.~\eqref{mainjpd} in all eigenvalues. Therefore, the joint probability density function of real and imaginary parts of \emph{unordered} eigenvalues of $\bH$ is given by
 \begin{align}
 \label{mainjpd_uo}
 \nonumber
\widehat{P}(\bet) &=\frac{1}{N!}\sum_{\{j\}}\frac{1}{[(2 \pi)^{2 N}\det\boldsymbol{\mathcal{T}}^{\{j\}}]^{1/2}} \\
&\times\exp\left[-\frac{1}{2} (\bet-\bnu^{\{j\}})^T (\boldsymbol{\mathcal{T}}^{\{j\}})^{-1} (\bet-\bnu^{\{j\}})\right],
\end{align}
where,
\begin{widetext}
    \begin{align*}
\bnu^{\{j\}}&=(\nu_{j_1},\nu_{j_2},\cdots,\nu_{j_{2N-1}},\nu_{j_{2N}} )^T, \hspace{0.8em}
\boldsymbol{\mathcal{T}}^{\{j\}}=\begin{pmatrix} 
\mathscr{T}_{j_1,j_1} & \mathscr{T}_{j_1,j_2} & \cdots & \mathscr{T}_{j_{1},j_{2N-1}} & \mathscr{T}_{{j_1},j_{2N}} \\
\mathscr{T}_{j_2,j_1} & \mathscr{T}_{j_2,j_2} & \cdots & \mathscr{T}_{j_{2},j_{2N-1}} & \mathscr{T}_{j_{2},j_{2N} }\\
\vdots & \vdots & \ddots & \vdots & \vdots \\
\mathscr{T}_{j_{2N-1},j_1} & \mathscr{T}_{j_{2N-1},j_2} & \cdots & \mathscr{T}_{j_{2N-1},j_{2N-1}} & \mathscr{T}_{j_{2N-1},j_{2N}} \\
\mathscr{T}_{j_{2N},j_1} & \mathscr{T}_{j_{2N},j_2} & \cdots & \mathscr{T}_{j_{2N},j_{2N-1}} & \mathscr{T}_{j_{2N},j_{2N} }
 \end{pmatrix},
\end{align*}
\end{widetext}
such that the sum in Eq.~\eqref{mainjpd_uo} involves the $N!$ permutations of index-pairs $\{(j_1,j_2), (j_3,j_4),...,(j_{2N-1},j_{2N})\}$ over the pairs $\{(1,2),(3,4),...,(2N-1,2N)\}$. From this, the marginal distributions can be obtained by integrating out the unwanted $\eta$-variables. For example, the unordered counterpart of Eq.~\eqref{md1}, i.e., the probability density of real part and the imaginary part of a generic eigenvalue of $\bH$ will be a sum of Gaussians, i.e.,
\begin{align}
\label{md1uo}
\widehat{p}_\mathrm{Re}(\eta)=\frac{1}{N}\sum_{j=1}^N\frac{1}{(2\pi \mathscr{T}_{2j-1,2j-1})^{1/2}} \exp\left[-\frac{(\eta-\nu_{2j-1})^2}{2\mathscr{T}_{2j-1,2j-1}}\right],
\end{align}
\begin{align}
\label{md2uo}
\widehat{p}_\mathrm{Im}(\eta)=\frac{1}{N}\sum_{j=1}^N\frac{1}{(2\pi \mathscr{T}_{2j,2j})^{1/2}} \exp\left[-\frac{(\eta-\nu_{2j})^2}{2\mathscr{T}_{2j,2j}}\right].
\end{align}

\begin{figure}[!t]
    \centering
    \includegraphics[width=0.9\linewidth]{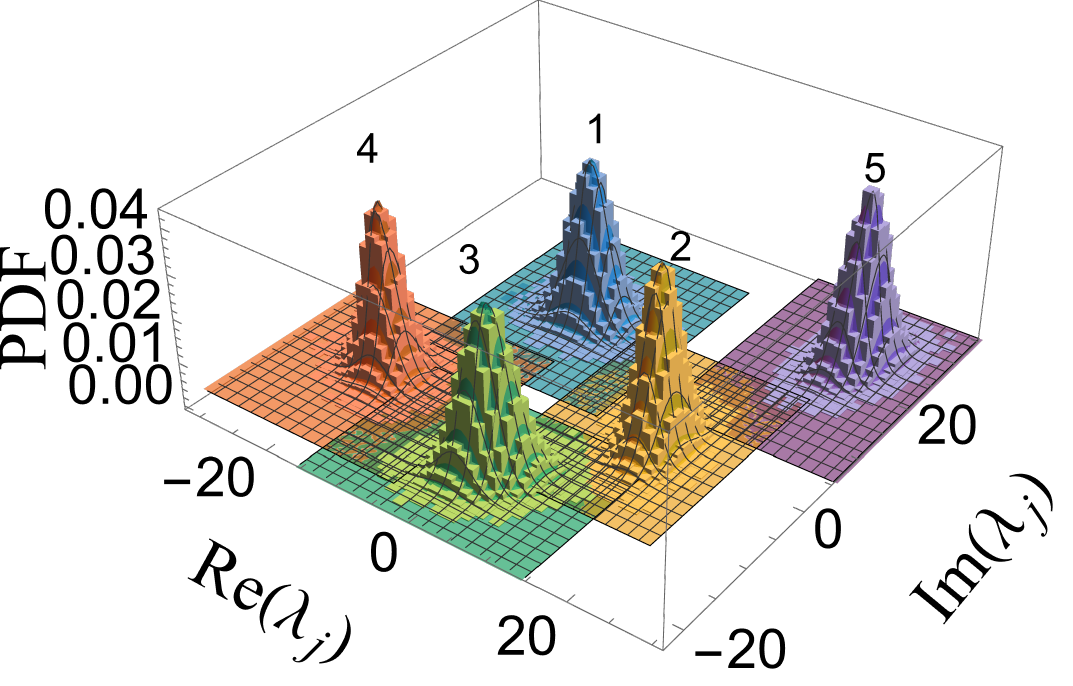}
    \caption{Probability densities of individual ordered eigenvalues of ${\bf H}$ in the complex plane for $N=5$. The averages and standard deviations of independent Gaussian elements of matrices ${\bf A}$ and ${\bf B}$ are $(u_1,u_2,u_3,u_4,u_5;\sigma_1,\sigma_2,\sigma_3,\sigma_4,\sigma_5)=(2,9,-7,-19/2,-5/3;1,2,1/2,2/7,4/5)$ and $(v_1,v_2,v_3,v_4,v_5;\tau_1,\tau_2,\tau_3,\tau_4,\tau_5)=(4,8,-15/2,3,20/3; 6/5, 2/3, 3/4,4/7,3/5)$, respectively. The simulation results, obtained from an ensemble comprising 100 000 matrices, are shown as histograms, while the two-dimensional surfaces are based on analytical result given in Eq.~\eqref{marginal-joint}. The ordering of the eigenvalues has been indicated using the numbers above the histograms.}
    \label{fig1}
\end{figure}

\begin{figure}[!t]
    \centering
    \includegraphics[width=0.98\linewidth]{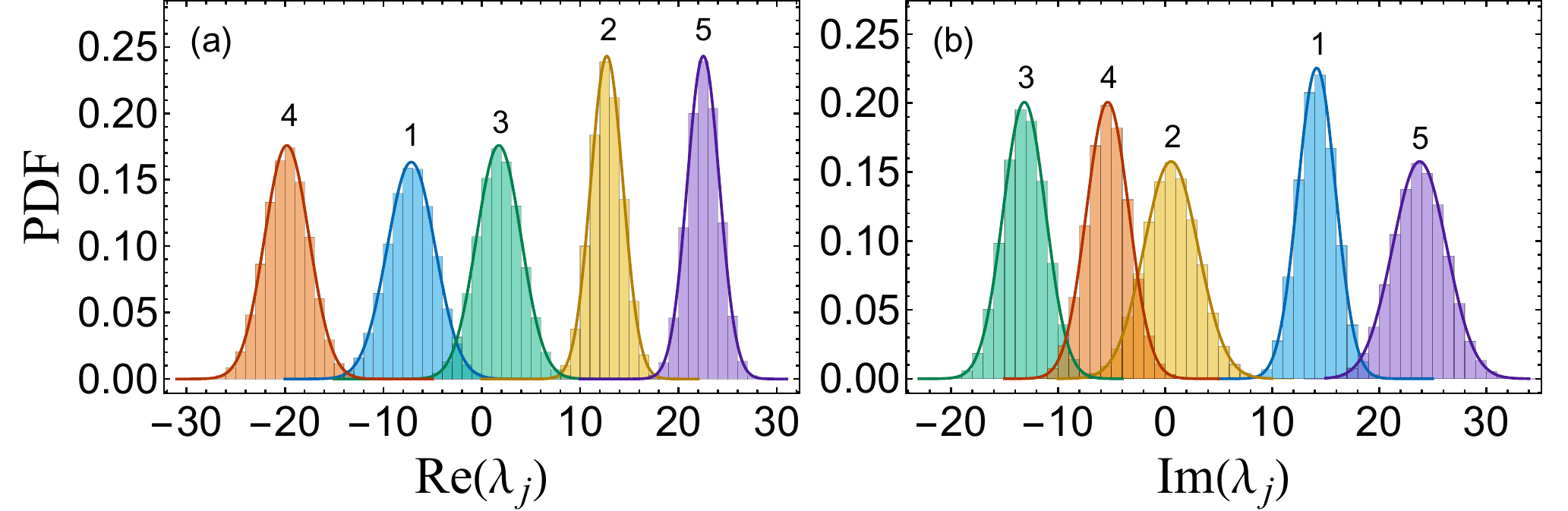}
    \caption{Probability densities of (a) real and (b) imaginary parts of individual ordered eigenvalues of ${\bf H}$. The parameter values are the same as those in Fig.~\ref{fig1}. The histograms depict the results obtained from numerical simulations, while the solid lines represent the analytical results. The numbers above the histograms indicate the ordering of the eigenvalues.}
    \label{fig2}
\end{figure}

\begin{figure}[!t]
    \centering
    \includegraphics[width=0.9\linewidth]{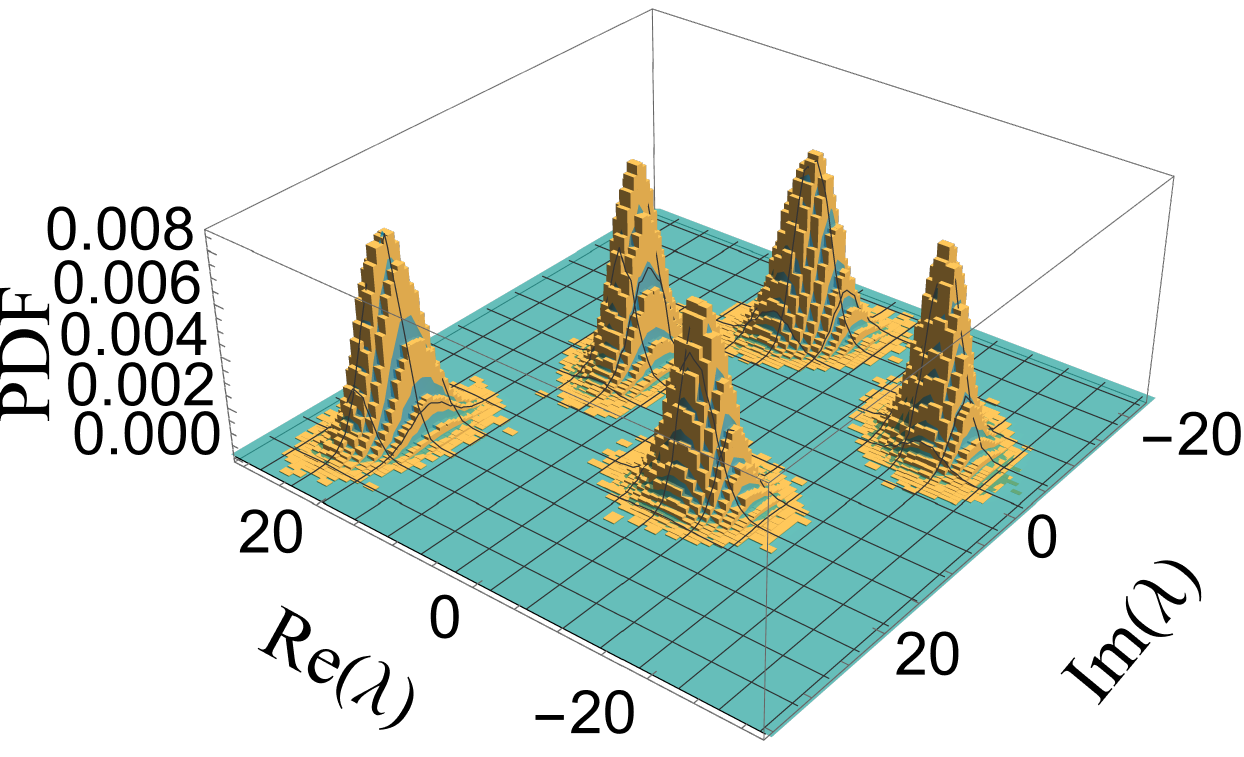
}
    \caption{Probability density of an unordered eigenvalue of ${\bf H}$ in the complex plane for $N=5$. Parameter values and presentation scheme are as in Fig.~\ref{fig1}. In this case, the analytical result employed is Eq.~\eqref{mainjpd_uo}.}
    \label{fig3}
\end{figure}

\begin{figure}[!t]
    \centering
    \includegraphics[width=0.98\linewidth]{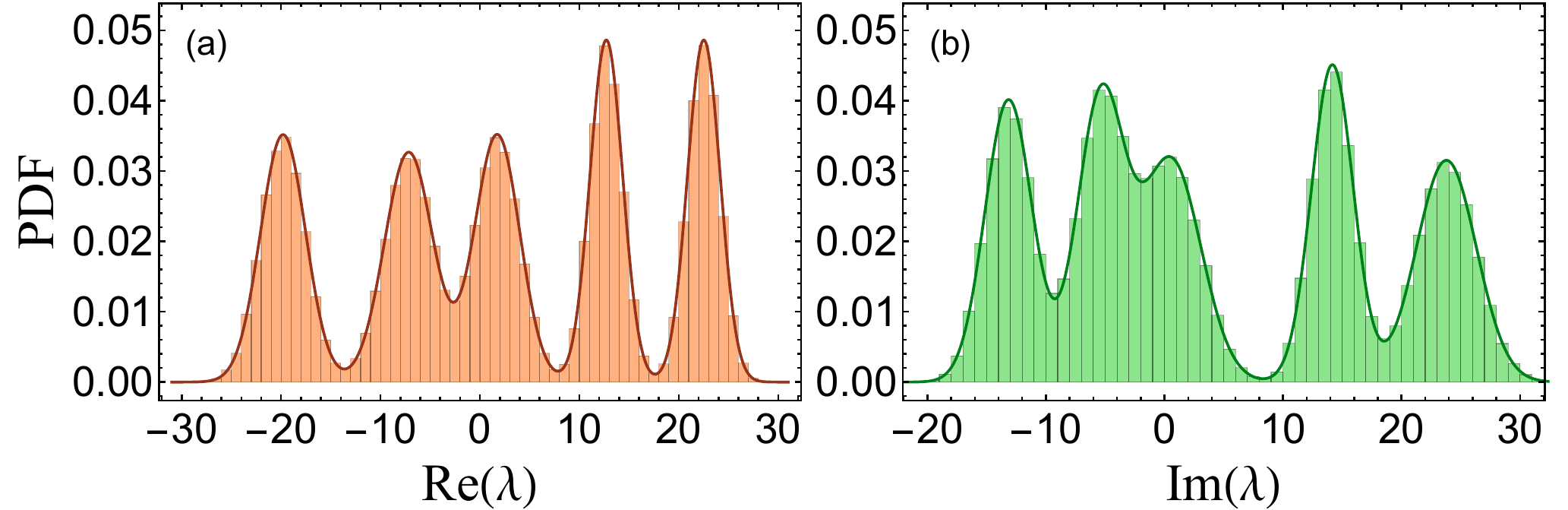}
    \caption{Probability densities of (a) real and (b) imaginary parts of an unordered eigenvalue of ${\bf H}$, corresponding to the one shown in Fig.~\ref{fig1}. The histograms are obtained from numerical simulations, while the solid lines are derived from Eqs.~\eqref{md1uo} and~\eqref{md2uo}.}
    \label{fig4}
\end{figure}

To provide a clearer comparison, we present the analytical results from this section alongside Monte Carlo simulations of the corresponding random matrix models. Firstly, we examine the eigenvalue density of ordered eigenvalues in the complex plane. Figure~\ref{fig1} illustrates an example using an ensemble of 100 000 ${\bf H}$ matrices of size $N = 5$, showing the probability density functions of individual eigenvalues as histograms in the complex plane. The figure caption specifies the corresponding parameter values. The numerical simulations, represented by the histograms, are compared to two-dimensional surfaces that depict the analytical results obtained from Eq.~\eqref{marginal-joint}. To further analyze the characteristics, we show the densities of the real and imaginary parts in Fig.~\ref{fig2}. Again, histograms obtained from simulations are used, while solid lines represent the analytical expressions. For the same parameter values, Figs.~\ref{fig3} and~\ref{fig4} present results for the unordered eigenvalue case, utilizing Eqs.~\eqref{mainjpd_uo} to \eqref{md2uo}.

Moreover, in Appendix, we examine the influence of tuning the means and variances of the matrix elements in $\bA$ and $\bB$ on the matrices $\bscrT$ and $\bnu$.


\section{Wigner-like matrix model based on $\bH$}
\label{SecWig}

The Wigner matrix model $(\bG+\bG^\dag)/2$, where $\bG$ is a square matrix with independent and identically distributed (iid) zero mean Gaussian elements (Ginibre random matrix) and $(\cdot)^\dag$ represents the conjugate transpose operation, is arguably the most popular one in the theory of random matrices~\cite{Mehta2004,Forrester2010}. It corresponds to the classic Gaussian ensembles, specifically Gaussian Orthogonal Ensemble (GOE) when $\bG$ is real and Gaussian Unitary Ensemble (GUE) when $\bG$ is complex. Wigner matrices are widely used to model complex, chaotic, and disordered systems across various fields. Originally introduced by Wigner in nuclear physics to study the energy level statistics of heavy nuclei~\cite{Wigner1955,Wigner1957,Wigner1958}, these matrices have since found applications in both physics and applied areas~\cite{GMW98,ABF18}. For large Wigner matrices, the eigenvalue density follows the well-known Wigner semi-circle law~\cite{Wigner1955,AGZ2009}, exhibiting a semi-circular (or, more precisely, semi-elliptical) shape, reflecting the universality that arises from the inherent symmetries of complex systems. However, as one moves away from the iid, zero mean set-up, deviations from the classic case are observed. We examine below this matrix model with $\bG$ replaced by the circulant matrix $\bH$ and study its eigenvalues.

Employing the eigenvalue decomposition $\bH=\bU^\dag \boldsymbol{\Lambda} \bU$, it is clear that we have
\begin{align}
\bR:=\frac{\bH+\bH^\dag}{2}=\bU^\dag\frac{(\boldsymbol{\Lambda}+\boldsymbol{\Lambda^\dag})}{2}\bU=\bU^\dag\mathrm{Re}(\boldsymbol{\Lambda})\bU,
\end{align}
which shows that the eigenvalues $\lambda^{R}_l$ of $\bR$ are just the real part of eigenvalues of $\bH$, i.e., $\lambda^{R}_l=\eta_{2l-1}$ for $l=1,...,N$. In a similar manner, we have,
\begin{align}
\bJ:=\frac{\bH-\bH^\dag}{2i}=\bU^\dag\frac{(\boldsymbol{\Lambda}-\boldsymbol{\Lambda^\dag})}{2i}\bU=\bU^\dag\mathrm{Im}(\boldsymbol{\Lambda})\bU.
\end{align}
Therefore, the eigenvalues $\lambda^{J}_l$ of $\bJ$ are the imaginary part of the eigenvalues of $\bH$, i.e., $\lambda^{J}_l=\eta_{2l}$ for $l=1,...,N$. In both cases, the joint distribution of these eigenvalues, as well as their marginals follow from the results in Sec.~\ref{SecMatMod}. These are completely different from the classical case, where the exact eigenvalue distributions are expressible in terms of weighted Hermite polynomials~\cite{Mehta2004, Forrester2010}.

\section{Wishart-like matrix model based on $\bH$}
\label{SecWis}
Another classical random matrix model is the Wishart model given by $\bG\bG^\dag$, where $\bG$ is a square Gaussian random matrix which in the general case, could be rectangular. For real $\bG$, one obtains the Laguerre Orthogonal Ensemble (LOE) whereas when $\bG$ is complex, the Laguerre Unitary Ensemble (LUE) is obtained. The study of Wishart matrices began in 1928 with the groundbreaking work of J. Wishart~\cite{Wishart1928} and has since found several applications in RMT particularly in the field of quantum information, quantum chaos, wireless communication, functional analysis, and so on~\cite{GMW98,ABF18}. The bulk behavior of the spectrum of large Wishart matrices is characterized by the Mar\v{c}enko-Pastur law~\cite{MP1967}, which defines their asymptotic spectral density. Here, we replace $\bG$ by the circulant matrix $\bH$ and analyse how the eigenvalue statistics deviate from the classical case. Thus, we have,
\begin{align}
\bW:=\bH \bH^\dag=(\bU^\dag \boldsymbol{\Lambda}\bU)(\bU^\dag\boldsymbol{\Lambda^\dag}\bU)=\bU^\dag|\boldsymbol{\Lambda}|^2\bU,
\end{align}
since $\bU\bU^\dag=\mathds{1}_N$. Therefore, the eigenvalues $\lambda^{W}_l$ of $\bW$ are modulus-squared eigenvalues of $\bH$, i.e., $\lambda^{W}_l=\eta_{2l-1}^2+\eta_{2l}^2$ for $l=1,...,N$. The joint probability density for these eigenvalues can be obtained using,
\begin{align}
P(\lambda^{W}_1,...,\lambda^{W}_N)=\int d\bet P(\bet)\prod_{l=1}^N \delta(\lambda^{W}_l-\eta_{2l-1}^2-\eta_{2l}^2).
\end{align}
Let us consider the corresponding multidimensional Laplace transform ($\{\lambda^{W}_l \mapsto s_l \}$),
\begin{align}
\bPs(s_1,...,s_N)=\int d\bet P(\bet)\prod_{l=1}^N e^{-s_l (\eta_{2l-1}^2+\eta_{2l}^2)}.
\end{align}
Defining
\begin{align*}
    \bS=\oplus_{l=1}^{N}(s_l\otimes\mathds{1}_2)=\mathrm{diag}(s_1,s_1,s_2,s_2,\ldots,s_N,s_N),
\end{align*}
and inserting the expression of $P(\bet)$ from Eq.~\eqref{mainjpd}, we obtain,
\begin{align}
\nonumber
&\bPs(s_1,...,s_N)=\frac{1}{[(2\pi)^{2N}\det \bscrT]^{1/2}}\\\nonumber
&~~~~~~~~~~~~~~~~~\times\int d\bet\exp[-\bet^T \bS \bet] \\&~~~~~~~~~~~~~~~~~\times\exp\left[-\tfrac{1}{2}(\bet-\bnu)^T\bscrT^{-1}(\bet-\bnu)\right].
\end{align}
This multidimensional Gaussian integral in ${\eta_i}$ can be performed to yield
\begin{align}
\bPs(s_1,...,s_N)=\frac{\exp\left[\frac{1}{2}\bnu^T \{(\bscrT+2\bscrT \bS \bscrT)^{-1}-\bscrT^{-1}\} \bnu \right]}{[\det(\mathds{1}_N+2 \bS \bscrT)]^{1/2}}.
\end{align}
Unfortunately, it does not seem feasible to perform the inverse Laplace transform to obtain the joint distribution of the eigenvalues. However, if we just focus on one of the eigenvalues (say $\lambda^W_j$) along with the assumption that the mean vector $\bmu$, and hence $\bnu$, is zero, then proceeding similar to above, we obtain the following expression for Laplace transform associated with the joint probability density of the real ($\eta_{2j-1}$) and imaginary ($\eta_{2j}$) parts,
\begin{align}
\psi(s)=\frac{1}{[\det(\mathds{1}_2+2 s \widetilde{\bscrT})]^{1/2}},
\end{align}
where $\widetilde{\bscrT}$ is as defined below Eq.~\eqref{marginal-joint}. Now, if 
\begin{align*}
t_{j}^{\pm} &=[(\mathscr{T}_{2j-1,2j-1}+\mathscr{T}_{2j,2j})\\ \nonumber &~~~\pm\sqrt{(\mathscr{T}_{2j-1,2j-1}+\mathscr{T}_{2j,2j})^2-4\mathscr{T}_{2j-1,2j}\mathscr{T}_{2j,2j-1}}]/2
\end{align*}
are the eigenvalues of $\widetilde{\bscrT}$, then the above can be written as 
\begin{align}
\psi(s)=\frac{1}{[(1+2st_{j}^{+})(1+2st_{j}^{-})]^{1/2}}.
\end{align}
The inverse Laplace transform~\cite{PBM1992} can then be performed to give,
\begin{align}
\nonumber
p_W(\lambda^{W}_j)&=\frac{1}{2(t_j^{+}t_j^{-})^{1/2}}\exp\left[-\left(\frac{1}{t_j^{+}}+\frac{1}{t_j^{-}}\right)\frac{\lambda^{W}_j}{4}\right] \\
&~~~~~\times I_0\left(\left| \frac{1}{t_j^{+}}-\frac{1}{t_j^{-}}\right|\frac{\lambda^{W}_j}{4}\right),
\end{align}
where $I_0(z)$ is the zeroth-order modified Bessel function of the first kind. If one examines the distribution of an eigenvalue without ordering, then the corresponding probability density function would be
\begin{align}
\label{circWisuo}
\widehat{p}_W(\lambda^{W})=\frac{1}{N}\sum_{j=1}^N p_W(\lambda^{W}_j).
\end{align}
Compared to the classical Wishart ensemble, where the eigenvalue density is expressible in terms of weighted Laguerre polynomials, this result is again very different.

\begin{figure}[!ht]
\includegraphics[width=0.8\linewidth]{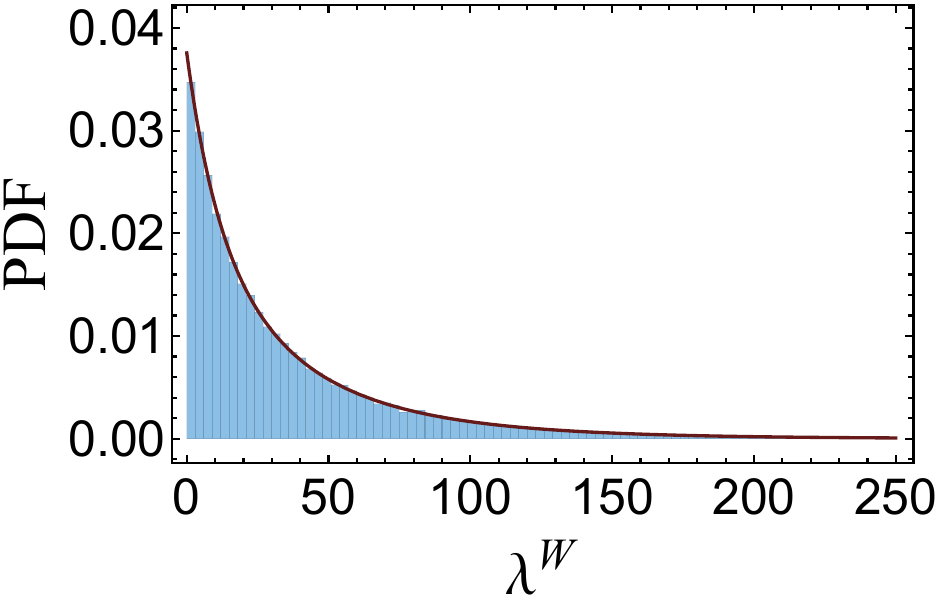}
\caption{
Probability density of an unordered eigenvalue of the matrix ${\bf W=HH^\dag}$ for $N=3$. The variances of independent zero-mean Gaussian elements of matrices ${\bf A}$ and ${\bf B}$ are $(\sigma_1,\sigma_2,\sigma_3)=(1,7/2,3/4)$ and $(\tau_1,\tau_2,\tau_3)=(4/3,2/3,9/2)$, respectively. The histogram has been obtained using numerical simulation comprising 20 000 matrices and the solid line is based on Eq.~\eqref{circWisuo}.
 }
    \label{fig5}
\end{figure}

As a validation, Fig.~\ref{fig5} showcases the distribution of a generic eigenvalue of the matrix $\bW$ for $N=3$, aligning with the analytical result in Eq.~\eqref{circWisuo}. Additional details are provided in the figure caption.

\section{Applications to random circulant graphs}
\label{Application}
In network science, circulant graphs are widely used for investigating structured and directional interactions in complex graphs. These are particularly valuable in scenarios where nodes exhibit regular and cyclic connectivity patterns, enabling the study of a wide range of network behaviors and processes~\cite{AYI2010,Stone1970,TLSPBK2016,EHH2021,EVBAWA2024,EFAR2013}. The cyclic nature of circulant graphs facilitates the exploration of phenomena such as network flow, system stability, and dynamic interactions, making them essential for optimizing the performance of various networked systems.

In this section, we explore the applications of our analytical results to model various random circulant graphs. We show that, despite the adjacency matrix elements in these graphs being non-Gaussian, their spectral statistics align well with those of the Gaussian circulant matrix model in the large-dimension limit. In the following subsections, we examine the spectral statistics of random circulant graphs with directed, undirected and double directed edges. These circulant graphs provide a novel framework for exploring network dynamics and offer valuable insights into the interaction between network properties and complex phases. Although primarily designed for unweighted adjacency matrices, the random circulant graph model can be considered a special case of the weighted version, and can also be explored in networks with complex weights, such as quantum and neural networks~\cite{BP2024,FMJBB2014,ZHBPH2021}. Furthermore, the spectral statistics of these models using RMT help us gain deeper insights into phenomena such as localization, community structure, randomness, and rigidity in complex networks, as highlighted in previous studies~\cite{JB2007,LZYSZ2006,PV2006,RJ2015}.

\subsection{Random circulant graph with directed edges}
\label{Application1}
Here, we focus on a random directed circulant graph model, which features circulant graphs with directed edges~\cite{ET1970}. In this model, the adjacency matrix ($\mathcal{A}$) of a directed graph is commonly represented with entries $\mathcal{A}_{jk}$ set to 1 to indicate a directed edge from node $j$ to node $k$, and 0 otherwise. The circulant structure is maintained by ensuring that each node has a consistent pattern of connections, meaning that every node has the same configuration of outgoing and incoming edges. A schematic of this structure is illustrated in Fig.~\ref{fig6}.

\begin{figure}[!ht]
\label{fig6}
  \begin{minipage}{0.5\textwidth}
    \centering
    \includegraphics[width=0.5\textwidth]{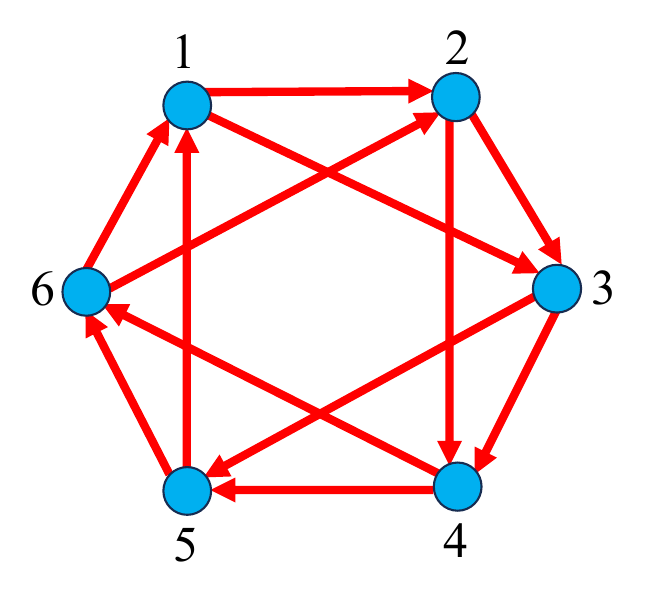}
  \end{minipage}
  \begin{minipage}{0.5\textwidth}
    \begin{eqnarray*}
  \centering
\begin{bmatrix}
 0 & 1 &1& 0&0&0\\
 0&0 & 1 &1& 0&0\\
 0&  0&0 & 1 &1& 0\\
 0&0&  0&0 & 1 &1\\
1& 0&0&  0&0 & 1 \\
1&1& 0&0&  0&0 \\
  \end{bmatrix}~~~~
 \end{eqnarray*}  
  \end{minipage}
  \captionsetup{width=0.9\columnwidth} 
  \caption{Schematic of a directed circulant graph and its adjacency matrix representation. This example graph features six nodes, with each node having four directed edges: two outgoing and two incoming.}
  \label{fig6}
\end{figure}
\begin{figure*}[!t]
    \centering
    \includegraphics[width=1.\linewidth]{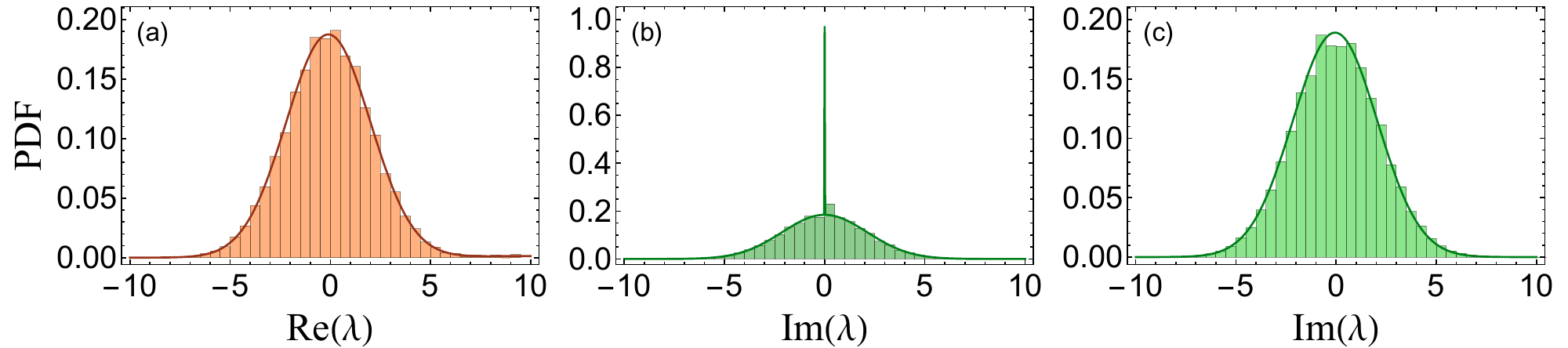}
    \caption{Probability densities of (a) real and [(b, c)] imaginary parts of a generic eigenvalue of adjacency matrix of random directed circulant graph and their comparison with analytical results for random circulant matrix model of Eq.~\eqref{matmod}. The histograms are based on simulation of 2000 adjacency matrices of dimension $N=100$, with edge probability $p_e=1/10$. Accordingly, the averages and variances of independent elements of matrix ${\bf A}$ are $(u_1, \sigma_1^2)= (0,0)$, $(u_2$ to $u_{100}, \sigma_2^2$ to $\sigma_{100}^2$)=(1/10, 9/100). For the matrix ${\bf B}$, we have considered two scenarios: $(v_1$ to $v_{100}, \tau_1^2$ to $\tau_{100}^2)=(0, 10^{-6})$, and $(v_1$ to $v_{100},\tau_1^2$ to $\tau_{100}^2)=(0, 0)$. The solid lines are based on these parameter values in our analytical results for Gaussian circulant matrix model. See the main text for details.}
    \label{fig7}
\end{figure*}

In our exploration of random directed circulant graphs, the edge probability $p_e$ governs the likelihood of a directed edge between any two nodes, thereby also serving as a parameter to characterize the sparsity of the adjacency matrix. By specifying the edge pattern for a single node, or equivalently a single row (or column) of the adjacency matrix, the circulant structure ensures that the patterns for the remaining nodes are automatically determined.

We compare our analytical results with numerical simulations of the random directed circulant graph with certain edge probability $p_e$. We focus on the first row of the adjacency matrix, as the other rows are permutations of this row. Apart from the first element, which is zero (indicating no self-loops and thus zero diagonal elements), the remaining entries are random, being either 1 or 0 with probabilities determined by $p_e$. Specifically, these entries are i.i.d. random variables from a Bernoulli distribution, with 1 occurring with probability $p_e$ and 0 with probability $1-p_e$~\cite{Uspensky1937}. Consequently, the mean and variance of these entries are $p_e$ and $p_e(1-p_e)$, respectively.

Anticipating identical spectral behavior in the large-dimension limit, in order to use our analytical results based on circulant matrix with Gaussian entries, we consider $\sigma_1\to0,u_1\to0$ so that $a_1=0$, and for the rest of the first row elements $a_2,...,a_N$ of $\bA$, we substitute these mean and variance values, i.e., we set $u_j=p_e$ and $\sigma_j^2=p_e(1-p_e)$. On the other hand, for the elements $b_1,...,b_N$ of $\bB$, we consider all the averages and variances approach zero. This setup corresponds to one of the cases discussed in Appendix, where each matrix realization has one purely real eigenvalue if $N$ is odd and two purely real eigenvalues if $N$ is even. In case we want to examine only the statistics of nonzero imaginary parts, along with real parts, we integrate out the Dirac-delta factor(s) that arise in the analytical joint PDF due to the above choices of mean and variance. Therefore, while the PDF of real part of a generic eigenvalue is given by Eq.~\eqref{md1uo}, for the PDF of imaginary part of a generic \emph{nonzero} eigenvalue, we use the following expressions for odd $N$ and even $N$, respectively, which exclude the \emph{zero} imaginary parts:
\begin{align}
\label{jpd_odd}
\widehat{p}_\mathrm{Im,O}(\eta)=\frac{1}{N-1}\sum_{j=2}^N\frac{  \exp\left[-\frac{(\eta-\nu_{2j})^2}{2\mathscr{T}_{2j,2j}}\right]}{(2\pi \mathscr{T}_{2j,2j})^{1/2}},
\end{align}
\begin{align}
\label{jpd_even}
\widehat{p}_\mathrm{Im,E}(\eta)=\frac{1}{N-2} \sum_{j=2 \atop (j\ne N/2+1)} ^{N}\frac{ \exp\left[-\frac{(\eta-\nu_{2j})^2}{2\mathscr{T}_{2j,2j}}\right]}{(2\pi \mathscr{T}_{2j,2j})^{1/2}} .
\end{align}
\begin{figure}[!t]
    \centering
    \includegraphics[width=.8 \linewidth]{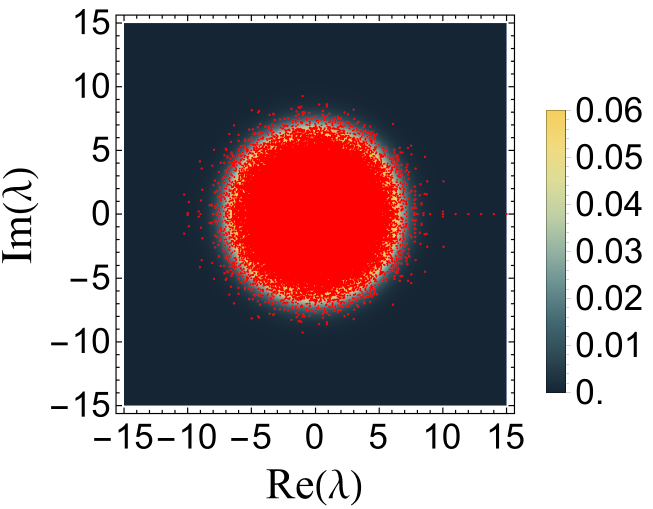}
    \caption{Scatter plot of eigenvalues obtained from the adjacency matrix of random directed circulant graph with parameters as in Fig.~\ref{fig7}. The background density plot is using Eq.~\eqref{mainjpd_uo} derived for the random circulant matrix model of Eq.~\eqref{matmod}.} 
    \label{fig8}
\end{figure}

In Fig.~\ref{fig7}, we present the numerically generated probability density functions for the real and imaginary parts of unordered eigenvalues of the adjacency matrix for a random directed circulant graph, and compare these with our analytical results as discussed above. We have considered $N=100$ and $p_e=1/10$ in generating the directed circulant graph. Therefore, in our analytical expressions, we set, $u_1=0, \sigma_1=0$, while for $j=2,...,N$, we use $u_j=p_e=1/10, \sigma_j^2=p_e(1-p_e)=9/100$. We also set $v_1,...,v_N=0$, but consider two scenarios for the $\tau_j$. In the first scenario, instead of setting $\tau_j$ exactly zero, we assign them a value $10^{-3}$. With these choices, the distribution for imaginary parts of $\lambda_1$ and $\lambda_{N/2+1}$ (for $N=100$ in this case) are not Dirac delta functions, but rather turn out to be sharply peaked Gaussian functions. In this case, we use Eqs.~\eqref{md1uo} and~\eqref{md2uo} for comparison with the directed circulant graph results. In the second scenario we set $\tau_j=0$ exactly, which makes the imaginary parts of $\lambda_1$ and $\lambda_{N/2+1}$ identically zero. In this case, we use Eq.~\eqref{md1uo} for the real part and Eq.~\eqref{jpd_even} for the imaginary part to compare with the nonzero imaginary parts of the eigenvalues of the adjacency matrix. In Fig.~\ref{fig7}(a), we show the probability density of the real eigenvalue which remains indistinguishable for the above two scenarios and also shows excellent agreement with Eq.~\eqref{md1uo}. In Fig.~\ref{fig7}(b), we examine the first scenario, where we observe a peak in the density at zero, corresponding to very small imaginary parts of $\lambda_1$ and $\lambda_{N/2+1}$. In Fig.~\ref{fig7}(c), we address the second scenario, focusing only on the nonzero imaginary parts and using Eq.~\eqref{jpd_even} for comparison. We find excellent agreement in all these plots.

In Fig.~\ref{fig8}, we present a scatter plot of eigenvalues in the complex plane from the directed graph adjacency matrix and compare it with the density plot derived from our analytical formula for the circulant random matrix. This comparison is based on the first scenario for the choice of $\tau_j$, as discussed above.

\subsection{Random circulant graph with undirected edges}
\label{Application2}

\begin{figure}[!t]
\begin{minipage}{0.5\textwidth}
\centering
\includegraphics[width=0.5\textwidth]{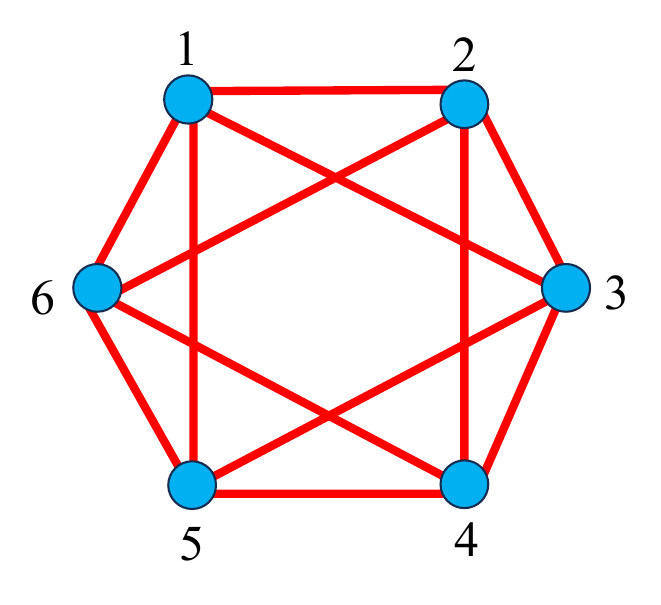}
\end{minipage}
\begin{minipage}{0.5\textwidth}
\begin{eqnarray*}
\centering
\begin{bmatrix}
 0 & 1 &1& 0&1&1\\
 1&0 & 1 &1& 0&1\\
 1&  1&0 & 1 &1& 0\\
 0&1&  1&0 & 1 &1\\
1& 0&1&  1&0 & 1 \\
1&1& 0&1&  1&0 \\
\end{bmatrix}~~~~
\end{eqnarray*} 
\end{minipage}
 \captionsetup{width=0.9\columnwidth} 
\caption{Schematic of an undirected circulant graph and its adjacency matrix representation. In this example, there are six nodes, each connected by four undirected edges.} 
\label{fig9}
\end{figure}

We now examine the spectrum of a random undirected circulant graph, also referred to simply as a circulant graph~\cite{ET1970,AP1979,MM1997,Vilfred2004,RR2018}. In an undirected circulant graph, each node exhibits uniform connectivity similar to that in a directed circulant graph, meaning every node shares the same pattern of connections with other nodes. The key difference is that in an undirected circulant graph, the connections are undirected rather than directed. An example of such a graph along with it adjacency matrix is shown in Fig.~\ref{fig9}. To generate these graphs randomly, we again consider edges present with a certain edge probability $p_e$.
\begin{figure}[!t]
\centering
\includegraphics[width=0.98\linewidth]{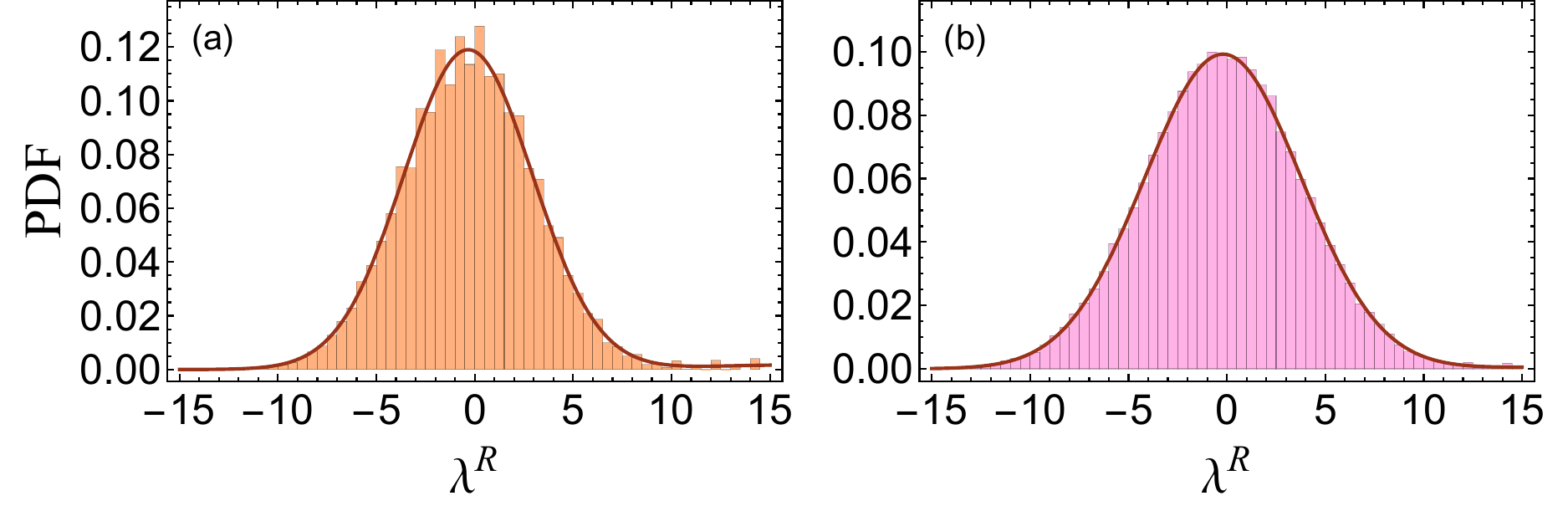}
\caption{Probability density of a generic eigenvalue of a random undirected circulant graph, compared with the analytical result for the Wigner-like matrix $\bR$ discussed in Sec~\ref{SecWig}. In panel (a), we have considered $N=50$ and $p_{e}=1/3$. The corresponding averages and variances of the matrix ${\bf A}$ elements are $(u_1, \sigma_1^2)= (0, 0)$ and $(u_2$ to $u_{50}, \sigma_2^2$ to $\sigma_{50}^2)=(1/3, 4/9)$, respectively. In panel (b), the parameters are $N=101$ and $p_{e}=1/5$, with averages and variances $(u_1, \sigma_1^2)= (0, 0)$ and $(u_2$ to $u_{101}, \sigma_2^2$ to $\sigma_{101}^2)=(1/5, 8/25)$, respectively. In both cases, for matrix elements of ${\bf B}$, the averages and variances have all been set to zero. The histograms in panels (a) and (b) are based on numerical simulations of 5000 and 2000 adjacency matrices, respectively, while the solid line follows Eq.~\eqref{md1uo}.}
\label{fig10}
\end{figure}
\begin{figure}[!ht]
\begin{minipage}{0.5\textwidth}
\centering
\includegraphics[width=0.5\textwidth]{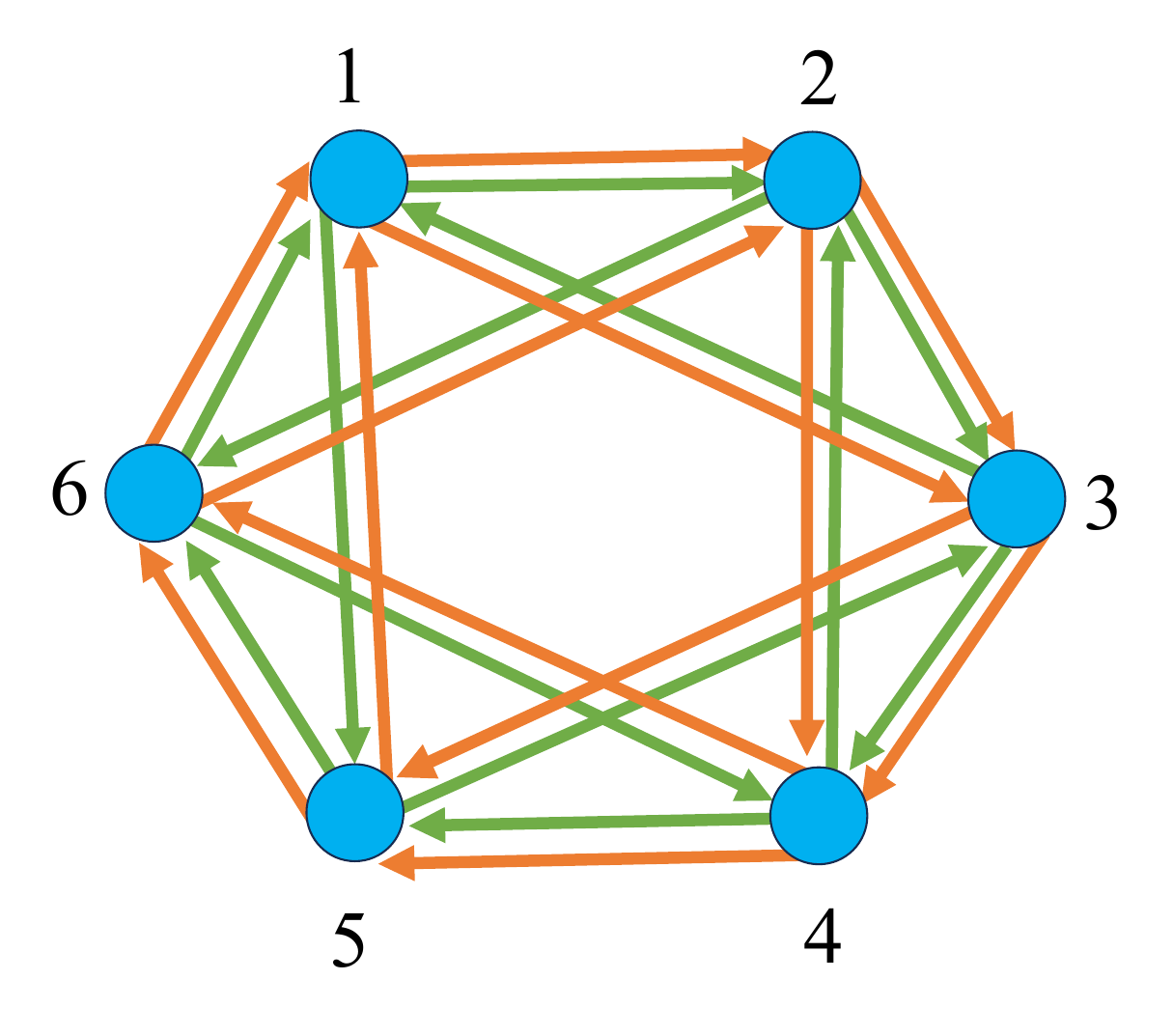}
\end{minipage}
\begin{minipage}{0.5\textwidth}
\begin{eqnarray*}
\centering
\begin{bmatrix}
 0 & 1 &1& 0&0&0\\
 0&0 & 1 &1& 0&0\\
 0&  0&0 & 1 &1& 0\\
 0&0&  0&0 & 1 &1\\
1& 0&0&  0&0 & 1 \\
1&1& 0&0&  0&0 \\
\end{bmatrix}+ \begin{bmatrix}
 0 & i &0& 0&i &0\\
 0&0 & i &0& 0&i\\
 i&  0&0 & i &0& 0\\
 0&i&  0&0 & i &0\\
0& 0&i&  0&0 & i \\
i &0& 0&i &  0&0 \\
\end{bmatrix}~~~~
\end{eqnarray*}  
\end{minipage}
 \captionsetup{width=0.9\columnwidth} 
\caption{Schematic diagram of a double-edged directed circulant graph and its adjacency matrix representation. The graph features six nodes, each connected by eight directed edges--four of one type and four of another, distinguished by orange and green colors.} 
\label{fig11}
\end{figure}
\begin{figure*}[!t]
    \centering
    \includegraphics[width= 1.\linewidth]{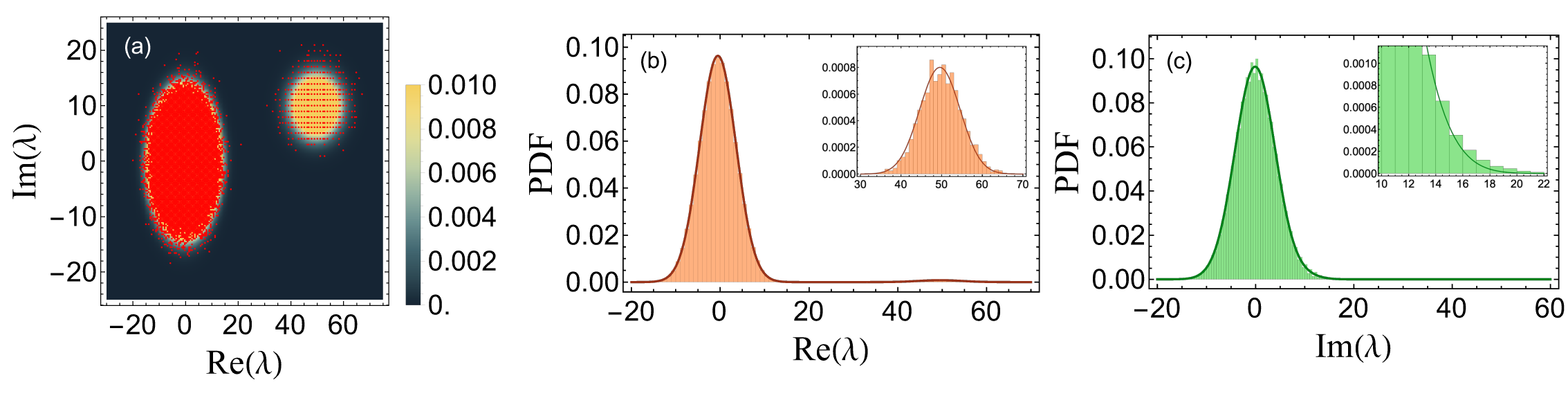}
    \caption{Scatter plot (a) and distribution of (b) real part, and (c) imaginary part of unordered eigenvalue of the complex adjacency matrix of random double-edged directed circulant graph for $N=100, p_e^{(1)}=1/2$ and $p_e^{(2)}=1/10$. These have been compared with the analytical results for the Gaussian circulant matrix $\bH$. For the random adjacency matrix, an ensemble comprising 3000 samples have been used. The averages and standard deviations of independent elements of the constituent matrices $\bA$ and $\bB$ of $\bH$ are $(u_1, \sigma_1^2)= (0, 0)$, $(u_2$ to $u_{100}$, $\sigma_2^2$ to $\sigma_{100}^2$)=(1/2,1/4), $(v_1, \tau_1^2)= (0,0)$, and $(v_2$ to $v_{100}$, $\tau_1^2$ to $\tau_{100}^2)=(1/10, 9/100)$. In panel (a), the background density plot is based on Eq.~\eqref{mainjpd_uo}. In panels (b) and (c) the solid lines are based on the analytical results in Eqs.~\eqref{md1uo} and~\eqref{md2uo}. The inset in (b) zooms in on the region around $\mathrm{Re}(\lambda) = 50$.}
\label{fig12}
\end{figure*}

\begin{figure*}[!t]
    \centering
    \includegraphics[width= 1.\linewidth]{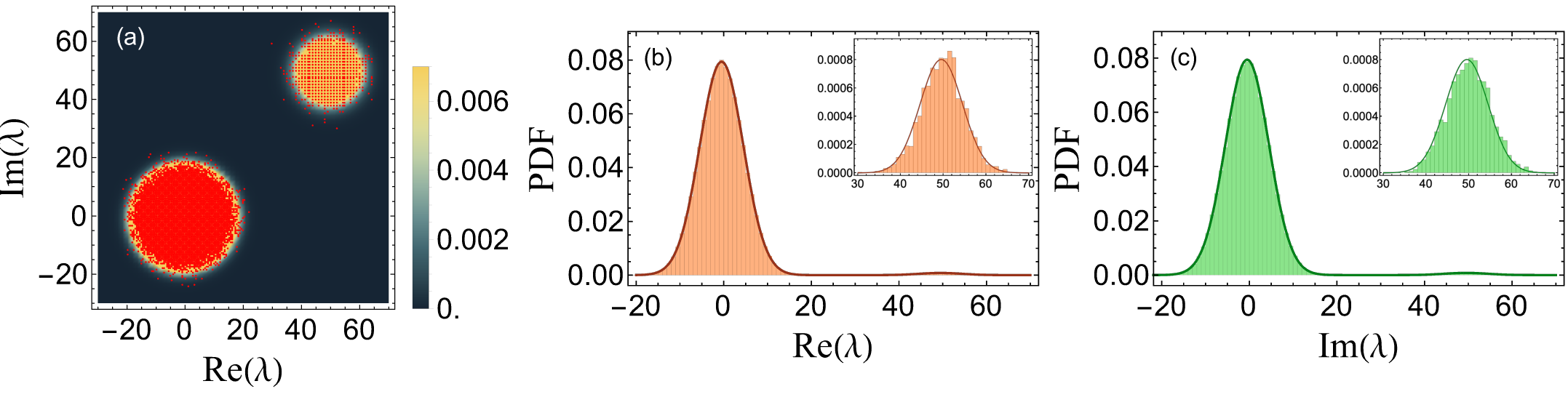}
    \caption{Plots as in Fig.~\ref{fig12}, now with parameters $N=100$ and $p_e^{(1)}=p_e^{(2)}=1/2$ for the complex adjacency matrix of a random double-edged directed circulant graph. The corresponding means and variances of the matrices $\bA$ and $\bB$ are $(u_1, \sigma_1^2) = (0, 0)$, $(u_2$ to $u_{100}$, $\sigma_2^2$ to $\sigma_{100}^2) = (1/2, 1/4)$, $(v_1, \tau_1^2) = (0, 0)$, and $(v_2$ to $v_{100}, \tau_2^2$ to $\tau_{100}^2) = (1/2, 1/4)$. The insets in (b) and (c) zooms in on the regions around $\mathrm{Re}(\lambda)=50$ and $\mathrm{Im}(\lambda)=50$, respectively. }
    \label{fig13}
\end{figure*}
In this case, since the adjacency matrix is symmetric, the appropriate matrix model for comparison is our Wigner-like matrix $\bR$ from Sec.~\ref{SecWig}. The number of independent elements in the adjacency matrix (excluding the diagonal zeros) is $[N/2]$, where $[\,\cdot\,]$ denotes the integer part. Furthermore, since we are dealing with real matrices, we need to consider the Wigner-like matrix with $\bB$ a zero matrix. To determine the parameters for the non-symmetric $\bA$ in our model, we note that if the mean and variance of its element $a_j$ are $u_j$ and $\sigma_j^2$, then the mean and variance of the corresponding element in $(\bA+\bA^T)/2$ would be $u_j$ and $\sigma_j^2/2$. Assuming that the spectral statistics of $(\bA+\bA^T)/2$ match those of the adjacency matrix of the random undirected circulant graph, we set $u_j=p_e$ and $\sigma_j^2/2=p_e(1-p_e)$ for $j=2,3,...,N$. For $a_1$, which lies on the diagonal, we set $u_1=0$ and $\sigma_1=0$. Additionally, for $\bB$ we set $v_j=0$ and $\tau_j=0$ for all $j$.

Figure~\ref{fig10} shows the spectral density derived from the adjacency matrix of a random undirected circulant graph for (a) $N=50$ and (b) $N=101$, based on ensembles of 5000 and 2000 matrices, respectively. The other parameters are specified in the figure caption. The results are compared with the analytical predictions for the eigenvalues of the Wigner-like matrix $\bR$ discussed in Sec.~\ref{SecWig}, demonstrating excellent agreement.

\subsection{Random circulant graph with directed double edges}
\label{Application3}

We finally apply our analytical results to a random double-edged directed circulant graph model.  In the conventional representation of the  adjacency matrix $(\mathcal{A})$ of a directed graph with multiple edges, one assigns $\mathcal{A}_{jk}$ a value equal to the number of edges directed from node $j$ to node $k$, and others zero~\cite{Harary1994}. However, in the case of a double-edged directed graph having distinct meanings to the two edges, we may represent it using a complex representation in the following way. If there is only first-type edge connected from node $j$ to $k$, we assign $\mathcal{A}_{jk}=1$, if there is only second-type edge connected from node $j$ to $k$, we assign $\mathcal{A}_{jk}=i$, and if both kinds of edges are present, we have $\mathcal{A}_{jk}=1+i$. Other elements are assigned a value of zero. An example of this representation is illustrated in Fig.~\ref{fig11}. As evident, the configuration of two types of edges can also be interpreted as a two-layer multiplex network~\cite{KT2006}. In this context, the complex representation offers an alternative to the traditional block structure of a two-layer multiplex network~\cite{RJ2022}, enabling both layers to be represented within a single block.

For constructing random double-edged directed circulant graph, we assign the two kinds of edges with edge probabilities, say $p_{e}^{(1)}$ and $p_{e}^{(2)}$. Following the approach outlined in Sec.~\ref{Application1}, for comparison with our general analytical results for the matrix $\bH=\bA+i\bB$, we set $u_1=0,\sigma_1^2=0,v_1=0,\tau_1^2=0$ for $a_1$ and $b_1$, For the remaining elements ($j=2,...,N$), we use $u_j=p_e^{(1)},\sigma_j^2=p_e^{(1)} (1-p_e^{(1)}) , v_j=p_e^{(2)}$, and $\tau_j^2=p_e^{(2)} (1-p_e^{(2)})$. 

In Fig.~\ref{fig12}, we compare our analytical results with the distribution of an unordered eigenvalue obtained from the adjacency matrix for $N=100, p_{e}^{(1)}=1/2$ and $p_e^{(2)}=1/10$. The number of adjacency matrices used in the ensemble is 3000. Similarly, Fig.~\ref{fig13} shows the results for $p_{e}^{(1)}=p_{e}^{(2)}=1/2$. In both cases, we find strong agreement between the results from the random adjacency matrix and our analytical results for the Gaussian circulant matrix model.

\section{Summary and conclusion}
\label{SecSum}
In this work, we considered a versatile random matrix model defined by ${\bf H}={\bf A} + i {\bf B}$, where ${\bf A}$ and ${\bf B}$ are real circulant matrices with independent but non-identical Gaussian entries. Through rigorous analytical calculations, we derive the exact joint probability density of this matrix model and demonstrate its multivariate Gaussian nature. This also enabled us to derive the marginal density functions of arbitrary order. By manipulating the averages and variances of the Gaussian elements, we showcased the model's ability to interpolate across a wide range of eigenvalue distributions in the complex plane, including those associated with the transition from random real circulant to complex circulant matrix. Additionally, we extended our investigation to include Wigner-like and Wishart-like matrices constructed from these random circulant matrices. Finally, we also demonstrated the application of our results for random circulant graphs for which the circulant adjacency matrices involve non-Gaussian elements.

Future research could delve into numerous  captivating directions. To illustrate, an avenue worthy of exploration involves the examination of circulant matrices-based variants of the elliptic Ginibre ensemble~\cite{BF2024,ADM2023,Molag2023,SK2024} and non-Hermitian Wishart matrices~\cite{ABK2021,BBD2023} that are encompassed by the traditional Gaussian random matrices. Moreover, exploring the implications of incorporating additional matrix structures or constraints within the present framework could lead to newer insights. Additionally, investigating non-Gaussian distributions for the matrix entries and studying their impact on the eigenvalue distributions would provide valuable insights into the robustness of this model. We have already taken some steps in this direction by applying our results to compare with the spectra of adjacency matrices for circulant graphs, which feature non-Gaussian elements.

\vspace{1em}
\textit{Note added.} While this manuscript was under review, one of the coauthors, Santosh Kumar, passed away.

\section*{Acknowledgments}
S.S. and H.S. acknowledges Shiv Nadar Institution of Eminence for financial support.  S.K. acknowledges the support provided by SERB, DST, Government of India, via Grant No. CRG/2022/001751. Last, we also thank Dr. Ayana Sarkar, Departmente de Physique and Institut Quantique, Université de Sherbrooke, and Dr. Aritra Laha of the Department of Physics, Shiv Nadar Institution of Eminence, for fruitful discussions and critical reading of the paper. We are deeply grateful to our coauthor, Santosh Kumar (now deceased), for his invaluable contributions to the conceptualization and development of this work. His insightful guidance and dedication greatly enriched this research.

\setcounter{equation}{0}
\renewcommand{\theequation}{A\arabic{equation}}

\section*{
Appendix: Explicit evaluations of $\bscrT$ and $\bnu$}
\label{AppenExpli}


In this section, we provide explicit expressions for $\bscrT$ and $\bnu$ in terms of the averages and variances of matrix elements of ${\bf A}$ and ${\bf B}$.
From the definition of $\bscrT=\bQ^T\Sigma \bQ$, and noting that $\bK_1^T=\bK_1, \bK_2^T=\bK_2$, it follows that the matrix elements of $\bscrT$ are given by
\begin{align}
\mathscr{T}_{2 l-1,2 m-1}&=\bt_l^T \bK_1 \bSg \bK_1 \bt_m=\sum\limits_{r=1}^{N}\left(\sigma_r^2 C_{l, r}C_{m, r}+\tau_r^2 S_{l, r}S_{m, r}\right),
\end{align}
\begin{align}
\mathscr{T}_{2 l-1,2 m}&=\bt_l^T \bK_1 \bSg \bK_2 \bt_m=\sum\limits_{r=1}^{N}\left(\sigma_r^2 C_{l, r}S_{m, r}-\tau_r^2 S_{l, r}C_{m, r}\right),
\end{align}
\begin{align}
\mathscr{T}_{2 l,2 m-1}&=\bt_l^T \bK_2\bSg \bK_1 \bt_m=\sum\limits_{r=1}^{N}\left(\sigma_r^2 S_{l, r}C_{m, r}-\tau_r^2 C_{l, r}S_{m, r}\right),
\end{align}
\begin{align}
\label{Tee}
\mathscr{T}_{2 l,2 m}&=\bt_l^T \bK_2\bSg \bK_2 \bt_m=\sum\limits_{r=1}^{N}\left(\sigma_r^2 S_{l, r}S_{m, r}+\tau_r^2 C_{l, r}C_{m, r}\right),
\end{align}
for $l,m=1,2,...,N$. Similarly, for the mean vector $\bnu$, we obtain
\begin{align}
\nu_{2l-1}=\bt_j^T \bK_1\bmu =\sum\limits_{r=1}^{N}\left[C_{l, r}u_r-S_{l, r}v_r\right],
\end{align}
\begin{align}
\nu_{2l}=\bt_j^T \bK_2\bmu=\sum\limits_{r=1}^{N}\left[C_{l, r}v_r+S_{l, r}u_r\right],
\end{align}
for $l=1,...,N$. Some special cases of the above deserve further discussion, as outlined below.

\subsection*{1. Some special cases of the covariance matrix $\bscrT$}\label{app:Subseccovar}

\begin{enumerate}[(i)]

\item
{\it $\sigma_r=\tau_r$ \text{for} $r=1,...,N$}

In this case
\begin{align}
\mathscr{T}_{2 l-1,2 m-1}=\sum_{r=1}^{N}\sigma_r^2\cos{\left[\frac{2\pi(l-m)(N-r+1)}{N}\right]},
\end{align}
\begin{align}
\mathscr{T}_{2 l-1,2 m}=-\sum_{r=1}^{N}\sigma_r^2\sin{\left[\frac{2\pi(l-m)(N-r+1)}{N}\right]},
\end{align}
\begin{align}
\mathscr{T}_{2 l,2 m-1}=\sum_{r=1}^{N}\sigma_r^2\sin{\left[\frac{2\pi(l-m)(N-r+1)}{N}\right]},
\end{align}
\begin{align}
\mathscr{T}_{2 l,2 m}=\sum_{r=1}^{N}\sigma_r^2\cos{\left[\frac{2\pi(l-m)(N-r+1)}{N}\right]},
\end{align}
for $l,m=1,2,...,N$.

\item
{\it $\sigma_r=\tau_r=\sigma$ \text{for} $r=1,...,N$}

In this case, we obtain
\begin{align}
\nonumber
\mathscr{T}_{2 l-1,2 m-1}&=\mathscr{T}_{2l,2m}\\
\nonumber
&=\sigma^2\sum_{r=1}^{N}\cos{\left[\frac{2\pi(l-m)(N-r+1)}{N}\right]}\\
&=\sigma^2 \sum_{r=1}^N \delta_{l,m}=N \sigma^2\delta_{l,m},
\end{align}
where $\delta_{l,m}$ is the the Kronecker delta. On the other hand, 
\begin{align}
\nonumber
\mathscr{T}_{2 l-1,2 m}&=-\mathscr{T}_{2l,2m-1}\\
\nonumber
&=-\sigma^2\sum_{r=1}^{N}\sin{\left[\frac{2\pi(l-m)(N-r+1)}{N}\right]}\\
&=0.
\end{align}
Therefore, overall, in this case we have
\begin{align}
\bscrT=N\sigma^2 \mathds{1}_{2N},
\end{align}
which means that the $\eta$'s become independent Gaussians.

\item
{\it $\tau_r\to 0$ \text{for} $r=1,...,N$}
\label{app:subsubsection}

In this case, the elements $\mathscr{T}_{2l,2l}$ warrant special attention. From Eq.~\eqref{Tee}, we find that they simplify to $\sum_{r=1}^N \sigma_r^2 S_{l,r}^2$. Now, $S_{l,r}=\sin[2\pi(l-1)(N-r+1)/N]$, which vanishes for all $r=1,...,N$ if $2(l-1)/N$ is an integer. This happens for $l=1$ and additionally when $l=N/2+1$ if $N$ is even. 

Working in a limiting sense, the consequence of this is that the variance $\mathscr{T}_{2,2}$ tends to zero in the Gaussian density (18) for $\eta_{2}$ and gives a Dirac delta function $\delta(\eta_{2}-\nu_2)$, thereby effectively making the imaginary part of the eigenvalue $\lambda_1$ of $\bH$ to assume a fixed value of $\nu_2$ for both even and odd $N$, i.e., $\mathrm{Im}(\lambda_1)\to \nu_2$. Additionally, in the even $N$ case, $\mathrm{Im}(\lambda_{N/2+1})\to \nu_{N+2}$. If on top of this, the corresponding averages $\nu$ (see below) are zero, then these eigenvalues become purely real. Therefore, a real circulant matrix will necessarily possess one real eigenvalue when $N$ is odd and two real eigenvalues when $N$ is even. The remaining eigenvalues occur in complex-conjugate pairs.

\item
{\it $\sigma_r\to 0$ \text{for} $r=1,...,N$}

This case is similar to the last one and now $\mathscr{T}_{2l-1,2l-1}$ becomes 0 when $l=1$ for both even and odd $N$, and additionally when $l=N/2+1$ for even $N$. Thus, here $\mathrm{Re}(\lambda_1)$ tends to a fixed value of $\nu_1$, i.e., $\mathrm{Re}(\lambda_1)\to \nu_1$. Moreover, when $N$ is even, $\mathrm{Re}(\lambda_{N/2+1})\to \nu_{N+1}$. Moreover, if the averages $\nu$ are zero , then these eigenvalues will be purely imaginary. The other eigenvalues occur in $\pm x+iy$ ($x,y\in \mathds{R}$) pairs. This is in consonance with the preceding case, as now we have a purely imaginary circulant matrix.

\end{enumerate}

\subsection*{2. Some special cases for the mean vector $\bnu$}\label{Subsecmean}

\begin{enumerate}

\item [(i)]
{\it $u_r=v_r=u$ \text{for} $r=1,...,N$}

In this case, we find that
\begin{align}
\bnu_{2 l-1}=\bnu_{2 l}=Nu\delta_{l,1}.
\end{align}
Consequently, we have
\begin{align}
\bnu=(Nu,Nu,0,\cdots,0)^T.
\end{align}

\item [(ii)]
{\it $u_r=u$ \text{and} $v_r=0$ \text{for} $r=1,...,N$}

In this case, we have
\begin{align}
\bnu_{2 l-1}=Nu\delta_{l,1},~ \bnu_{2 l}=0,
\end{align}
for $l=1,...,N$.

\item[(iii)]
{\it $u_r=0$ \text{and} $v_r=v$ \text{for} $r=1,...,N$}

In this case, we have
\begin{align}
\bnu_{2 l-1}=0, \bnu_{2 l}=N v \delta_{l,1},
\end{align}
for $l=1,...,N$. 

\end{enumerate}

\begin{figure*}[t!]
    \centering
    \includegraphics[width=\linewidth]{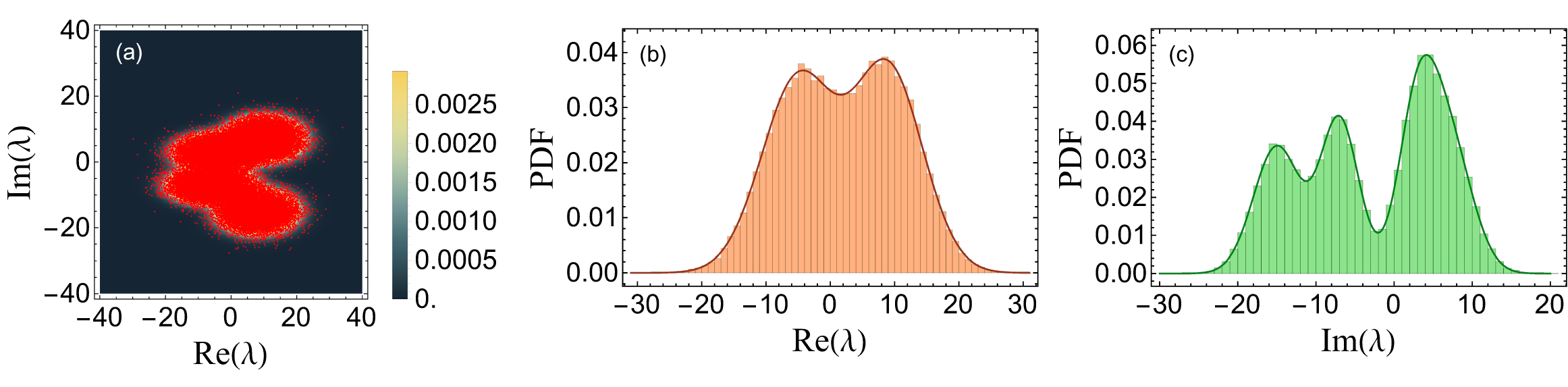}
    \caption{Distribution of an unordered eigenvalues of ${\bf H}$ for $N=4$. The averages and standard deviations of independent Gaussian elements of matrices ${\bf A}$ and ${\bf B}$ are $(u_1,u_2,u_3,u_4;\sigma_1,\sigma_2,\sigma_3,\sigma_4)=(2,6,-7,-5;5,2,1/2,4/3)$ and $(v_1,v_2,v_3,v_4;\tau_1,\tau_2,\tau_3,\tau_4)=(-3,2,1,3;7/4,3/2,1/4,5/6)$, respectively. Panel (a) shows the scatter plot of eigenvalues obtained from numerical simulation of 20000 matrices and the background density plot is using Eq.~\eqref{mainjpd_uo}. Panels (b) and (c) illustrate the probability densities of the corresponding real and imaginary parts. The histograms are based on the numerical simulation results, while the solid lines represent the densities calculated using Eqs.~\eqref{md1uo} and~\eqref{md2uo}.}
    \label{fig14}
\end{figure*}

\begin{figure*}[!t]
    \centering
    \includegraphics[width=\linewidth]{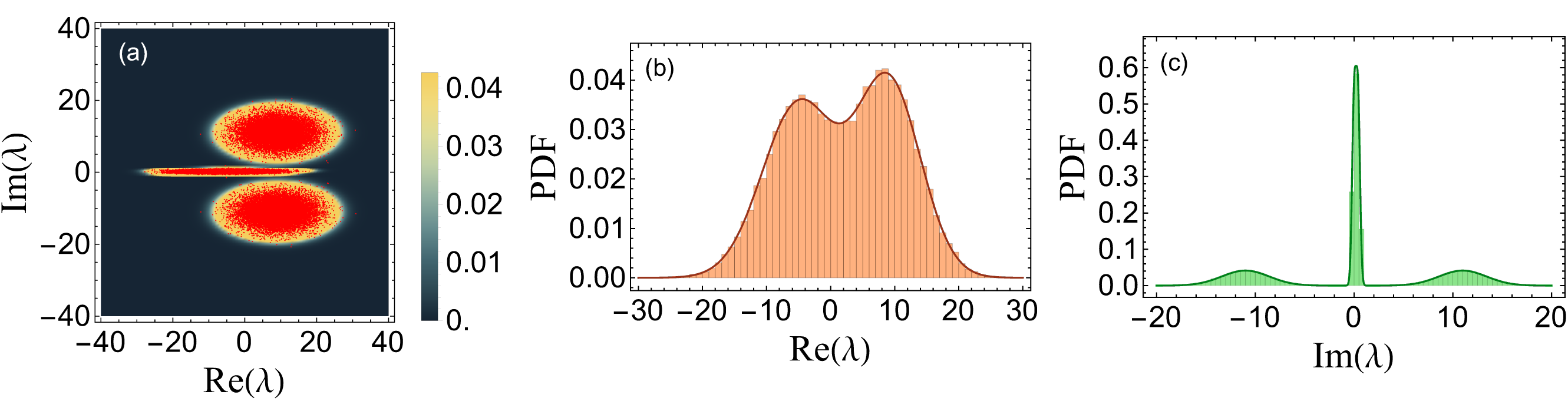}
    \caption{Plots as in Fig.~\ref{fig14} with all parameters same except those associated with matrix ${\bf B}$, given by $v_j=\tau_j=1/10$, $j=1,2,3,4$, in this case. The impact of making the averages and variances of matrix ${\bf B}$ close to zero is clearly seen. In consonance with the discussion~(\ref{app:subsubsection}) of Appendix 1, two eigenvalues approach the real line, while the other two tend to form complex conjugate pairs, leading to the results depicted above.}
    \label{fig15}
\end{figure*}

We examine the impact of changing the means $(v_j)$ and variances $(\tau_j^2)$ of matrix elements in matrix ${\bf B}$ on the eigenvalue density in Figs.~\ref{fig14} and~\ref{fig15} by employing 20000 ${\bf H}$ matrices of dimension $N=4$. By bringing these values close to zero in the second of these two figures (see captions), two eigenvalues approach the real line, while the other two tend to form complex conjugate pairs, as discussed in case~(\ref{app:subsubsection}) of Appendix 1. The first panel in both of these figures show the scatter plot of numerically generated eigenvalues in the complex plane along with density plot based on Eq.~\eqref{mainjpd_uo}. Additionally, the probability densities of the real and imaginary parts of the unordered eigenvalues are included in these figures.

\end{document}